\documentclass{article}
\usepackage{url}

\usepackage[margin=1in]{geometry}  
\usepackage{graphicx} 
\usepackage[hidelinks]{hyperref}
\usepackage{longtable}
\usepackage{tabu}
\title{Advancing Malicious Website Identification: A Machine Learning Approach Using Granular Feature Analysis}
\author{Kinh Tran\\
University of Guelph, Guelph, Ontario, Canada. \\
Email: kinh@uoguelph.ca
\and
Dusan Sovilj \\ Arctic Wolf Networks, Waterloo, Ontario, Canada. \\
Email: dusan.sovilj@arcticwolf.com}

\begin{document}

\maketitle

\begin{abstract}
Malicious website detection is an increasingly relevant yet intricate task that requires the consideration of a vast amount of fine details. Our objective is to create a machine learning model that is trained on as many of these finer details as time will allow us to classify a website as benign or malicious. If malicious, the model will classify the role it plays (phishing, spam, malware hosting, etc.). We proposed 77 features and created a dataset of 441,701 samples spanning 9 website classifications to train our model. We grouped the proposed features into feature subsets based on the time and resources required to compute these features and the performance changes with the inclusion of each subset to the model. We found that the performance of the best performing model increased as more feature subsets were introduced. In the end, our best performing model was able to classify websites into 1 of 9 classifications with a 95.89\% accuracy score. We then investigated how well the features we proposed ranked in importance and detail the top 10 most relevant features according to our models. 2 of our URL embedding features were found to be the most relevant by our best performing model, with content-based features representing half of the top 10 spots. The rest of the list was populated with singular features from different feature categories including: a host feature, a robots.txt feature, a lexical feature, and a passive domain name system feature.
\end{abstract}

\section{Introduction}
\label{sec:intro}
 Billions of people rely on the Internet daily to manage finances, consume news, interact with others, and accomplish tasks \cite{datareportal2024internet}. Due to the to the universal ease of access in creating and hosting a website on the Internet, the online space has become a monolithic network of websites that provide various services such as banking, social media, video games, or e-commerce \cite{2}. However, not all websites are created for fair trade, legitimate use, or even legal activity. The tremendous size, and open nature of the Internet has made it an attractive target for criminal activity. The annual amount of damage caused by reported cybercrime has increased exponentially since 2018 \cite{IC3_FBI_2024}. Cybercriminals create malicious websites to victimize Internet by deceptive, harmful, and illegal means. Phishing websites trick users into voluntarily disclosing their private information. Command and control servers can orchestrate cyber attacks to spread malware or deny service. Spam websites send unsolicited and unwanted advertisements. Malware hosting websites distribute malware that can remotely control, deny service or steal information \cite{a2}. Malicious advertisement hosting websites host advertisements that can execute code on the browser to steal sensitive information. Host scanning websites passively collect as much personally-identifying information from the user to sell for a profit. Exploit kits host software and information to enable criminals to exploit other users for personal gain. Credit card skimmers collect credit card details from their victims to commit transaction fraud \cite{4}. The prevalence of malicious websites has risen with the popularity of the Internet over its lifespan. Tools were developed to detect and respond to the growing threat of malicious websites. The tools forced malicious websites to adapt to evade detection, which obligated the tools to improve accordingly. To keep up with the sophistication of modern malicious websites, researchers studied the potential of leveraging machine learning (ML) to quickly and accurately classify websites as benign or malicious. For example, a study by Chaiban et al. investigated various web and web intelligence features to determine which ones had the greatest importance when using ML and whether a website is malicious. We aim to expand upon the work by Chaiban et al. \cite{app12062806}. This study attempts to study the efficacy of malicious website detection ML models when trained on a dataset with features that capture a holistic state of the web and web threat intelligence. However, it misses many features that are important to consider when making a classification. Some features such as the presence of invisible button elements to trick users into triggering unwanted events, or passive domain name system information containing historical data that can indicate the reputation of the website are missing \cite{8780783, 10.1145/3387168.3387236}. Another limitation of the model proposed by Chaiban et al. is that their data only trains the ML model to make a binary judgement of malicious or benign. There are various types of malicious activities a website can perform. Different malicious activities require unique responses to adequately mitigate. Therefore, binary classifications provide an insufficient amount of detail for practical defensive measures to be taken. For example, a malicious website can be identified as a phishing website would require a different countermeasure than a command and control server \cite{ibm_incidence_response}. Knowing this, we aim to increase the efficacy of the model further with additional features that provide a more holistic state of the web as well as web threat intelligence. We aim to increase its utility with additional labels for granular malicious classifications and study how that affects the efficacy of the model. To accomplish this, we have curated our own dataset of 441,701 samples, 9 labels, and 75 features to contribute\cite{me}. The key findings that we found were that increasing the number of features to capture a more holistic state of web and web threat intelligence increased its accuracy. We also found that the model prioritizes features differently when detecting varying granular classifications for malicious websites. This study details how we were able to train a machine learning model that can detect 8 different kinds of malicious websites with a 95.89\% accuracy rate using our own constructed dataset. This paper is organized as follows. 
 
 Section \ref{sec:related_work} outlines the work and results found in the investigation done by Chaiban et al., and discusses the work that has been published since and their findings. Section \ref{sec:dataset} outlines dataset, explaining its features, sizes, labels, and an explanation about how it was built. Section \ref{sec:methodology} discusses experiments conducted and their motivations. Section \ref{sec:experimental_results} presents and examines the results of those experiments and discusses the insights gained. Section \ref{sec:conclusion} provides concluding remarks as well as where we would like to see the future direction of this research proceed towards.

\section{Related Work}
\label{sec:related_work}
We are expanding upon Investigating the Influence of Feature Sources for Malicious Website Detection by Chaiban et al. The researchers conducted a study to determine how important various web, and web intelligence features were in detecting malicious websites. They built their own dataset of 105,485 samples, with 61,080 labeled as benign and 44,405 as malicious with 48 features. The study found that among the most important features, URL embeddings consistently outperformed the other features. The study also concluded that image embeddings did not have a significant impact on the model with their dataset and the content features were irrelevant as well \cite{app12062806}. We will be expanding upon this work in 3 different ways. First, we will increase the number of samples in our dataset to be roughly 400\% times greater than theirs. Second we will be adding 50\% more features totaling 77 features to investigate the impact on model accuracy, and feature contribution rankings. Lastly, we will study the impact of splitting the malicious label into 8 granular labels of classification on the model \cite{a1}.

The study by Chaiban et al., was heavily influenced by A.K. Singh for his extensive dataset. Although they ultimately decided to not use the dataset, many of the features found in A.K. Singh's dataset can be found in their investigation \cite{app12062806, SINGH2020106304, 10.1007/978-3-319-50472-8_17}. Our investigation also uses many of the features proposed by A.K. Singh as we have preserved most of the features used in the investigation by Chaiban et al \cite{me}. 

Since the publication of the investigation by Chaiban et al., there have been several other papers that were published in this field. A study by Alsaedi et al. which analyzed the performance of cyber threat intelligence-based (CTI) features extracted from Google searches on publicly extracted data from Kaggle. The dataset contains 651,191 samples with 4 labels: benign, defacement, phishing, and malware link \cite{kaggle}. They found a significant performance increase when adding CTI features to their model over leveraging only URL-based features. Using CTI features along with URL-based features let their model achieve a 96.80\% accuracy \cite{s22093373}. While the data was promising, we wanted to collect features beyond only URL and threat intelligence features that could not be extracted from the dataset available on Kaggle. So we chose to use another source to collect uniform resource locators (URLs) and collect the features ourselves. Another study by Aljabri et al. assessed the performance of lexical, content and network features using the dataset provided by A.K. Singh \cite{app12062806}. The study investigates the use of deep learning models on the dataset but found that it performed worse than their Naïve Bayes model \cite{https://doi.org/10.1155/2022/3241216}. 
A study by McGahagan IV et al. studied the effectiveness of features found through leveraging unsupervised learning and compared it to a priori features. They concluded that the features discovered by their unsupervised learning model performed more efficiently and nearly as well as selecting features proposed from researchers \cite{MCGAHAGAN2021102374}. Another study by Naim et al. also attempts to reduce the reliance on researcher proposed features through a proposed framework \cite{naim2023malicious}. These studies contrast our work as all of our features are a priori. We investigate the potential improvements made by adding more features to provide a holistic state of the web to discover the most relevant features. \\

\section{Dataset}
\label{sec:dataset}
In order to curate a dataset, we focused on the features in Table \ref{tab:all_features} \cite{me}. Custom scripts were created to extract all of the features. The raw URLs and labels were obtained through querying various threat intelligence platforms that will be outlined later. The dataset can be found on GitHub.\footnote{\url{https://github.com/CyberScienceLab/Malicious_benign_websites/}} The Raw HTML was collected through web scraping tools created using Beautiful Soup \cite{beautiful_soup}. The host name and domain name were obtained through parsing the URL using tldextract and urllib3 \cite{tldextract, urllib3}. Custom scripts were then used to extract the rest of the proposed features through using the raw ones. Lexical features were extracted through parsing the URL. Content features were collected through parsing the raw HTML and web scraping then parsing any URLs found within the raw HTML. Image related content features were not collected due to the security limitations of our cloud infrastructure. Host features were found through performing various look-ups, such as WHOIS or IP addresses associated with the raw domain and URL. Embedding features calculated using Distilbert and Longformer transformers \cite{distilbert, longformer}. The other embeddings were not collected due to time and security limitations of our cloud infrastructure. Passive DNS information was collected by querying the threat intelligence platforms: IBM X-Force, and Open Threat Exchange using the raw domain and URL \cite{IBM_X-force, alienvault}.

\clearpage
\setlength\LTleft{-0.5cm}
\makeatletter
\g@addto@macro\tabu@setup{\def\tabu@aligndefault{x}}
\makeatother
\renewcommand{\arraystretch}{1.5}
\begin{tiny}
\begin{longtabu} to \textwidth {p{1cm} p{3cm} p{8cm} }
\caption{List of features covered in this study. The labels indicate: * part of the Chaiban et al.'s dataset, ° not part of our final dataset. } 
\label{tab:all_features} \\
\hline
Category&  Feature& Description\\
\hline
Raw&  URL of the website *, ° & String format of the URL of the website.\\
&  HTML page content *, ° & The HTML page code of the homepage of the website in string format.\\
&  Hostname *, ° & The extracted hostname of the website in string format.\\
&  Domain name °& The extracted domain name of each website in string format.\\
&  Label of the website * & A binary label that states whether the website is malicious or benign.\\
\hline
Lexical&  Length of the URL *	& The number of characters in the URL.\\
&  Number of underscores, semicolons, subdomains, zeros, spaces, hyphens, @ symbols, queries, ampersands, and equal signs *	& The number of each of these characters counted in the URL. Note that all these characters are considered as separate features.\\
&  Hostname length *	& The number of characters in the hostname extracted from the given website.\\
&  Ratio of digits to URL *& The number of digits divided by the length of the URL.\\
& Ratio of digits to hostname *&The number of digits divided by length of hostname.\\
& IP address in URL *	&A binary label that states if an IP address exists in the URL.\\
& Existence of @ symbol in URL *	&A label that states if the URL contains any @ characters.\\
& Domain length &The number of characters in the extracted domain name. \\
& Unique URL characters, numbers and letters	&Taking the total number of unique characters, numbers and letters in a URL. Note that each of these is their own feature. \\
& Ratio of letters to chars *	&The unique letter count divided by the total unique character count in a URL. \\
& Ratio of numbers to chars * &Count of numbers in a URL divided by the total unique character in a URL. \\
&Top-level domain * &The top-level domain of a website as a category label. \\
& Domain entropy & The entropy of the domain calculated using the Shannon Entropy formula. \\
& Whether the URL contains a portable executable extension ° & A binary label that states whether or not the URL contains extensions found at the end of portable executables such as ".exe", ".dll", etc.  \\
& Number of top-level domains in the URL & The number of top-level domains present in the subdomains, second-level domain, and suffix of the URL. \\
& URL entropy & The entropy of the full URL calculated using the Shannon Entropy formula.  \\
\hline
Content& Length of JavaScript code * &	Total number of characters of JavaScript code present in HTML content.\\
&Length of deobfuscated JS code *, ° 	&Length of the JavaScript code put through deobfuscation tool. \\
& URL count in content *	&The total number of URLs found in the HTML page content. \\
& Unique URL count in content *	& The unique number of URLs found in the HTML page content.\\
& Suspicious JS function count *	&The number of suspicious functions found in the JavaScript code. \\
& JavaScript function count *	&The total number of JavaScript functions called in the code. \\
& Number of browser function calls in the JavaScript code & The number of JavaScript function calls that belong to the browser object. \\
& Number of DOM function calls in the JavaScript code & The number of functions that directly interacts with the document object model tree \cite{mozilla_dom_api}. \\
& Number of browser function calls in external JavaScript files & The number of JavaScript function calls that belong to the browser object. \\
& Number of DOM function calls in external JavaScript files& The number of functions that directly interacts with the document object model tree. \\
& Length of external JavaScript files & Total number of characters of present in linked JavaScript files. \\
& Number of functions in the external JavaScript files& The total number of functions called in the code of linked JavaScript files. \\
& Number of suspicious function calls in the external JavaScript files& The number of suspicious functions found in linked JavaScript files.\\
& Content length *	&The total number of characters of the HTML content page. \\
& Script tag references	* &A count of the total number of script tags in the HTML page content. \\
& Contains HEX * &Binary indicator about webpage’s HTML content containing any hexadecimal characters. \\
& HEX length *	&The number of Hexadecimal characters in the HTML content page. \\
& Number of out of domain image sources in the HTML & The number of img tags in the HTML content whose source is from a another domain. \\
& Number of advertisements on the website ° & The number of references to adblock filtering rules found in the HTML content. The filter rules used are found in easylist \cite{easylist}. \\
& DCD MPEG-7 *, °	&The 5 dominant colors of an image. They are extracted as 5 different columns. \\
& Average length of JS arrays *	&The average length of arrays in the JS code present in HTML content.\\
& Maximum length of JS arrays *	&The maximum length of arrays in the JS code present in HTML content. \\
& Maximum array length in external Javascript files & The average length of arrays in the code present in sourced JavaScript files.\\
& Average array length in external JavaScript files & The maximum length of arrays in the code present in sourced JavaScript files. \\
& Length CSS &The total number of characters in HTML styles present in HTML content. \\
& Number of hidden CSS elements & The number of HTML styles in the HTML content used to hide an HTML element from the user. \\
& Length of external CSS files &The total number of characters in HTML styles present in linked CSS files. \\
& Number of hidden CSS elements in external CSS files& The number of HTML styles in linked CSS files used to hide an HTML element from the user.\\
& Existence of a robots.txt file & A binary label that states if appending robots.txt to the domain name will successfully resolve. \\
& Length of the robots.txt file & The total number of characters in the robots.txt web page.\\
& Number of disallow rules, allow rules, user agents, comments, and sitemaps & The number of each of these entries counted in the robots.txt web page. Note that all these characters are considered as separate features. \\
& Whether the robots.txt file disallows the root& A binary label that states whether the domain consents to web crawlers to scraping its root. The robots.txt file can disallow the root with the line "/". \\
\hline
Host	&IP address *	&String format of the IP address of the website.\\
& Geographic location *	&The name of the country the website is hosted in. \\ 
& WHOIS information *	&A binary label that states whether the WHOIS information for the website is complete or incomplete. \\
& HTTPS *	&A binary label that states whether the website is secure HTTP protocol or not. \\
& Is in Alexa’s top 1 million *, °	&A binary label that describes if the given domain name of a URL exists in Alexa’s top 1 million domains. \\
& Top-level domain register price & The full cost to register domain name with a particular top-level domain in USD. \\
& Top-level domain renew price & The full cost to renew domain name with a particular top-level domain in USD. \\
& Top-level domain transfer price & The full cost to transfer domain name with a particular top-level domain in USD. \\ 
& Top-level domain ICANN fee & The transaction-based fees that are paid to ICANN annually \cite{ICANN_fee}.  \\
& Domain Registrar &The registrar company used to get the domain. \\
& Number of IP addresses that resolve to the domain & The number of IP addresses mapped to various servers, load balancers and content delivery networks (CDNs). \\
\hline
Embedding	&Image embeddings *, °	&The embeddings of the website’s image. \\
& Content embeddings *, °	&The embeddings of the HTML page content. \\
& URL embeddings (Distilbert) *	&URL embeddings extracted using the Distilbert Tranformer. \\
& URL embeddings (Longformer) *	&URL embeddings extracted using the Longformer Tranformer. \\
& Mean statistic for embeddings *	&URL and image embeddings, as well as chi-squared feature selected embedding dimensions, are summarized as a single representative number—the mean of the embedding vector (or parts of the vector). \\
\hline
Passive DNS & Length of the Passive DNS history associated with the domain & The number of entries in a domain's passive DNS history. \\
& Number of unique IP addresses in the passive DNS & The number of unique IP addresses in a domain's passive DNS history. \\
& Number of unique hostnames in the passive DNS& The number of unique hostnames in a domain's passive DNS history. \\
& Number of countries in the passive DNS& The number of unique countries in a domain's passive DNS history. \\
& Number of suspicious ASNs in the passive DNS& The number of suspicious autonomous system names in a domain's passive DNS history. \\
& Number of false positive ASNs in the passive DNS& The number of false positive autonomous system names in a domain's passive DNS history.\\
& Number of ASN switches& The number of times a domain was stated to have switched autonomous system names by its passive DNS history.\\
\hline
\end{longtabu}
\end{tiny}

In addition to the newly proposed features, we wanted to improve upon the study conducted by Chaiban et al. \cite{app12062806}, by increasing the number of samples and dividing the malicious label into more granular classifications. We searched for sources of malicious URLs with granular labels to aggregate. From our search, we found IBM X-Force Exchange: an threat intelligence sharing platform that provides application programming interfaces (APIs) to query URLs based on granular classifications. Examples of classifications include: spam URLs, botnet command and control server, and phishing URLs \cite{IBM_X-force}. We also used another threat intelligence sharing platform called Open Threat Exchange. Their community sourced repository of indicators of compromise (IoC), known as AlienVault, provides malicious URLs with granular labels. Examples include: host scanners, exploit kits, and malicious advertisement hosting \cite{alienvault}. Lastly we rounded off the dataset with URLs from URLHaus and ThreatFox: two platforms by abuse.ch that host curated lists of malicious URLs with granular classifications. Examples include: malware hosting, and command and control servers \cite{urlhaus, threatfox}. The granular labels were not consistent between threat intelligence platforms in name but they were in definition. For example, some sources used the label "botnet", while others used "c2 server". We decided to combine these labels under command and control server \cite{a3,a4}. Even though these labels represent different things, they require each other to perform an attack in the context of malicious websites. The resulting dataset contains 441,701 samples: 235,721 benign, 73,345 phishing, 66,490 command and control server, 46,009 spam, 16,726 malware hosting, 3085 malicious advertisement hosting, 231 host scanners, 82 exploit kits, and 12 credit card skimmers. This dataset is approximately 400\% larger than the dataset built by Chaiban et al. with 9 potential labels vs 2 labels from Chaiban et al. Once the dataset was constructed and all the features were obtained. We considered a different approach to performing cascade style additions from the one conducted by Chaiban et al. In their investigation, they grouped the features into the following categories: lexical, host-based, content-based, url embeddings (Longformer), image embeddings, content embeddings, and url embeddings (Distilbert)\cite{app12062806}. However, because we wanted the groups to be based on time and resources required to compute, we grouped the features into the following 5 subsets: base features, cascade 1, cascade 2, cascade 3, and all features. The base features subset contains only features related to the URL and domain name of the website. Cascade 1 contains all of the features in the base features followed by features related to the robots.txt file that may or may not be present on the website. Cascade 2 contains cascade 1 along with features related to the websites HTML, JavaScript and CSS contents. Cascade 3 contains cascade 2 with passive DNS features. Lastly, all features contains cascade 3 and all of the host-based features\cite{a3}. 
\section{Methodology}
\label{sec:methodology}
The methodology consists of four experiments: a preliminary experiment to test the efficacy of the proposed features against the features in the study by Chaiban et al. \cite{app12062806}, a granular classification experiment to test the efficacy of the features when making a classifications beyond benign and malicious, and a performance optimization experiment to test how accurate the model is once the hyperparameters are tuned. The first experiments were performed using logistic regression and random forest classification with 10-folds cross validation using mostly default hyperparameters to calculate the scoring metrics: accuracy, AUC score, F1 score, Matthews coefficient, and precision. For random forest, we set the maximum number of features equal to the total number of features. The granular classification experiments were performed by repeating the same experiments previously mentioned but we replaced the malicious label with more granular classifications and stratified 10-folds cross validation to ensure that each label would be present in all told folds. We reduced our focus to just the test accuracy scores for each of the cascades. We decided to also add top 2 and top 3 accuracy scores to measure how well the model could predict the label if we included its top 2 and top 3 classifications respectively. Finally, we concluded the experiments by testing various hyperparameter optimizations to determine the highest accuracy we could achieve using XGBoost. We then introduced the URL embedding features by using Distilbert and Longformer transformers. For both transformers, we used linear discriminant analysis (LDA) and Chi-squared feature selection reductions and analyzed the impacts of their additions to our model. Linear discriminant analysis was chosen over the principal component analysis method used by Chaiban et al. \cite{app12062806} as it is more appropriate for supervised learning models \cite{linear_discriminant_analysis, jolliffe2002pca}.
\section{Experimental Results}
\label{sec:experimental_results}

\begin{table}[h]
    \tiny
    \centering
    \caption{Mean and Standard Deviation of Scoring Metrics Computed using 10-Fold Cross Validated Logistic Regression with only Benign and Malicious Labels}
    \vspace{5pt}
    \hspace*{-1cm} 
    \begin{tabular}{lcccccc}
        \hline
         &  Accuracy&  ROC AUC&   F1& Matthews Coefficient& Precision&Recall\\
        \hline
         Base&  0.7736 ± 0.0022&  0.8539  ± 0.0017&   0.7313 ± 0.0028& 0.5489 ± 0.0042& 0.8189 ± 0.0026&0.6607 ± 0.0042\\
         1&  0.7952 ± 0.001&  0.8813 ± 0.0015&   0.7696 ± 0.0012& 0.5883 ±  0.0021& 0.8093 ± 0.0025&0.7337 ± 0.0022\\
         2&  0.8668 ± 0.0013&  0.9398 ± 0.0001&   0.8589 ± 0.0016& 0.733 ± 0.0027& 0.8483 ± 0.0026&0.8699 ± 0.0025\\
         3&  0.8750 ± 0.0012&  0.9458 ± 0.0009&   0.868 ± 0.0014 & 0.7497 ± 0.0025& 0.8555 ± 0.0028&0.8808 ± 0.0019\\
         All&  0.8812 ± 0.0013&  0.949 ± 0.0008&   0.8749 ± 0.0016& 0.7632 ± 0.0026& 0.8635 ± 0.0026&0.8867 ± 0.0028\\
         \hline
    \end{tabular}
    \label{tab:logistic_regression_scoring_metrics}
\end{table}

\begin{table}[h]
    \tiny
    \centering
    \caption{Mean and Standard Deviation of Scoring Metrics Computed using 10-Fold Cross Validated Random Forest Classification with only Benign and Malicious Labels} 
    \vspace{5pt}
    \hspace*{-1cm} 
    \begin{tabular}{lcccccc}
        \hline
         &  Accuracy&  ROC AUC&   F1& Matthews Coefficient& Precision&Recall\\
         \hline
         Base& 0.9841 ± 0.0005 & 0.9974 ± 0.003 & 0.9829 ± 0.0005 & 0.968 ± 0.0010 & 0.9834 ± 0.0006 & 0.9824 ± 0.0012\\
         1& 0.9767 ± 0.0011 & 0.9963 ± 0.0004 & 0.9749 ± 0.0011 & 0.9532 ± 0.0021 & 0.9792 ± 0.0011 & 0.9707 ± 0.0016\\
         2& 0.9753 ± 0.001 & 0.997 ± 0.0002 & 0.9736 ± 0.0011 & 0.9504 ± 0.0021 & 0.9703 ± 0.0015 & 0.9769 ± 0.0011\\
         3& 0.9723 ± 0.0008 & 0.9966 ± 0.0002 & 0.9705 & 0.9444 ± 00016 & 0.9639 ± 0.0013 & 0.97713 ± 0.0009\\
         All& 0.9729 ± 0.0009 & 0.9965 ± 0.0003 & 0.9712 ± 0.0010 & 0.9457 ± 0.0019 & 0.9636 ± 0.0017 & 0.9789 ± 0.0008\\
         \hline
    \end{tabular}
    \label{tab:random_forest_scoring_metrics}
\end{table}

\subsection{Preliminary Experiments}
In the logistic regression experiment, we found that, on average, as the number of features increased, the aforementioned scoring metrics decreased as shown in Table \ref{tab:logistic_regression_scoring_metrics}. In the random forest experiment, we found that the scoring metrics increased until cascade 2 but then started to decrease past that as shown in Table \ref{tab:random_forest_scoring_metrics}. 

\begin{figure}
    \centering
    \includegraphics[width=0.7\textwidth]{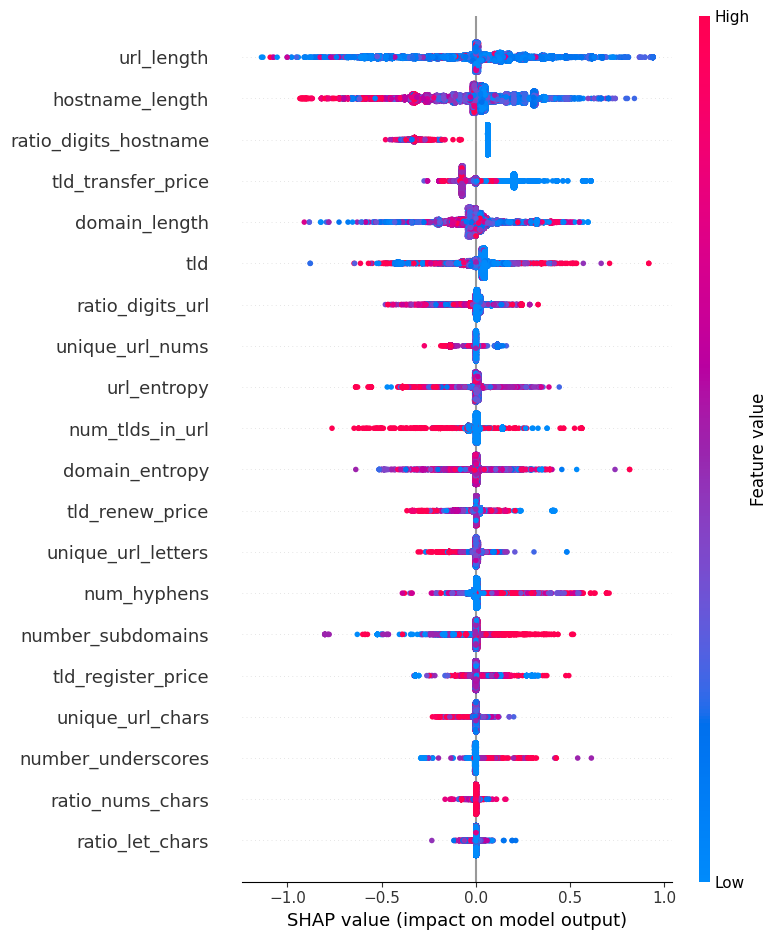} 
    \caption{SHAP Summary Graph of Impact on Random Forest Output vs Feature Values to the Rank Feature Importance of Base Features when making a Benign Classification on the First of Ten Stratified Folds}
    \label{fig:random_forest_feature_rankings_base}
\end{figure}

\begin{figure}
    \centering
    \includegraphics[width=0.7\textwidth]{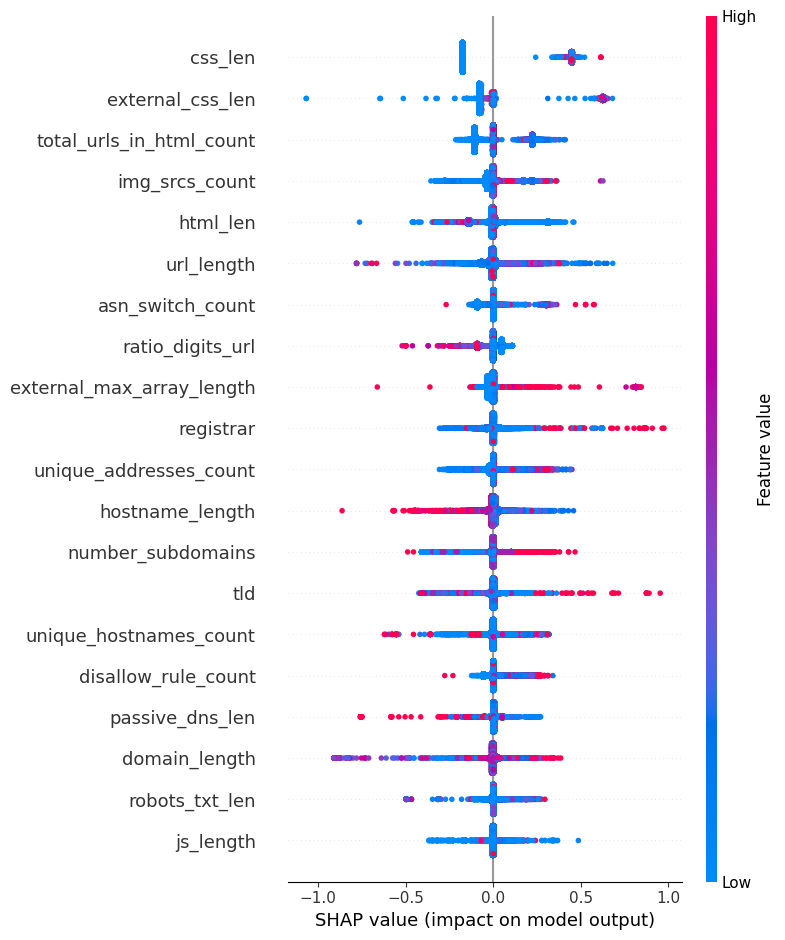} 
    \caption{SHAP Summary Graph of Impact on Random Forest Output vs Feature Values to the Rank Feature Importance of All Features when making a Benign Classification on the First of Ten Stratified Folds}
    \label{fig:random_forest_feature_rankings_all}
\end{figure}

\begin{table}
    \scriptsize
    \centering
    \caption{Mean and Standard Deviation of Scoring Metrics Computed using 10-Fold Cross Validated Logistic Regression and Random Forest Classification with Benign and Granular Malicious Labels}
    \vspace{5pt}
    \hspace*{-1cm} 
    \begin{tabular}{lccc}
        \hline
         Logistic Regression&  Accuracy&   Top 2 Accuracy&Top 3 Accuracy\\
         \hline
         Base& 0.7209 ± 0.0019&  0.8678 ± 0.0012&0.9373 ± 0.0009\\
         1& 0.7236 ± 0.0013&  0.8728 ± 0.001&0.9413 ± 0.001\\
         2& 0.7577 ± 0.0023&  0.8907 ± 0.0012&0.9474 ± 0.0012\\
         3& 0.7721 ± 0.0019&  0.8967 ± 0.001&0.9513 ± 0.0009\\
         All& 0.7766 ± 0.0021&  0.8988 ± 0.0014&0.954 ± 0.001\\
         \hline
         Random Forest Classifier& Accuracy&  Top 3 Accuracy&Top 2 Accuracy\\
         \hline
         Base& 0.8692 ± 0.0014&  0.9409 ± 0.0007&0.9727 ± 0.0003\\
         1& 0.8823 ± 0.0014&  0.9495 ± 0.0011&0.9755 ± 0.0004\\
         2& 0.9219 ± 0.0014&  0.9688 ± 0.0005&0.9857 ± 0.0004\\
         3& 0.9206 ± 0.0011&  0.9676 ± 0.0006&0.9852 ± 0.0006\\
         All& 0.928 ± 0.0016&  0.9714 ± 0.0009&0.9876 ± 0.0003\\
         \hline
    \end{tabular}
    \label{tab:granular_scoring_metrics}
\end{table}

\begin{figure}
    \centering
    \includegraphics[width=0.7\textwidth]{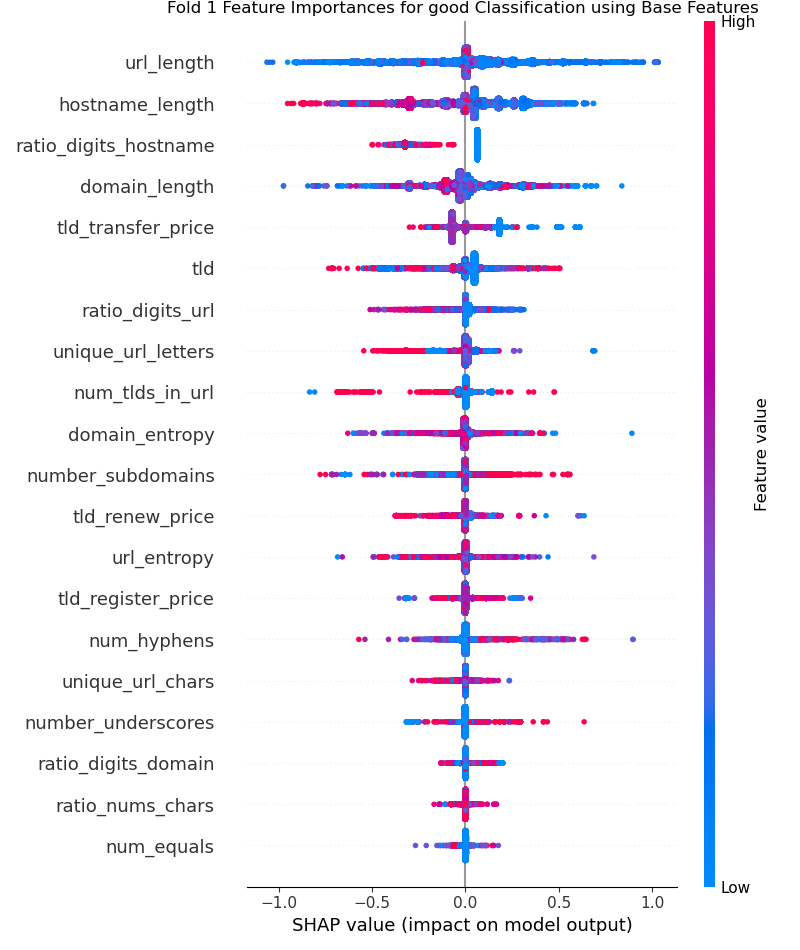} 
    \caption{SHAP Summary Graph of Impact on Random Forest Classifier Output vs Feature Values to the Rank Feature Importance of Base Features when making a Benign Classification on the First of Ten Stratified Folds}
    \label{fig:random_forest_feature_rankings_good_base}
\end{figure}

\begin{figure}
    \centering
    \includegraphics[width=0.7\textwidth]{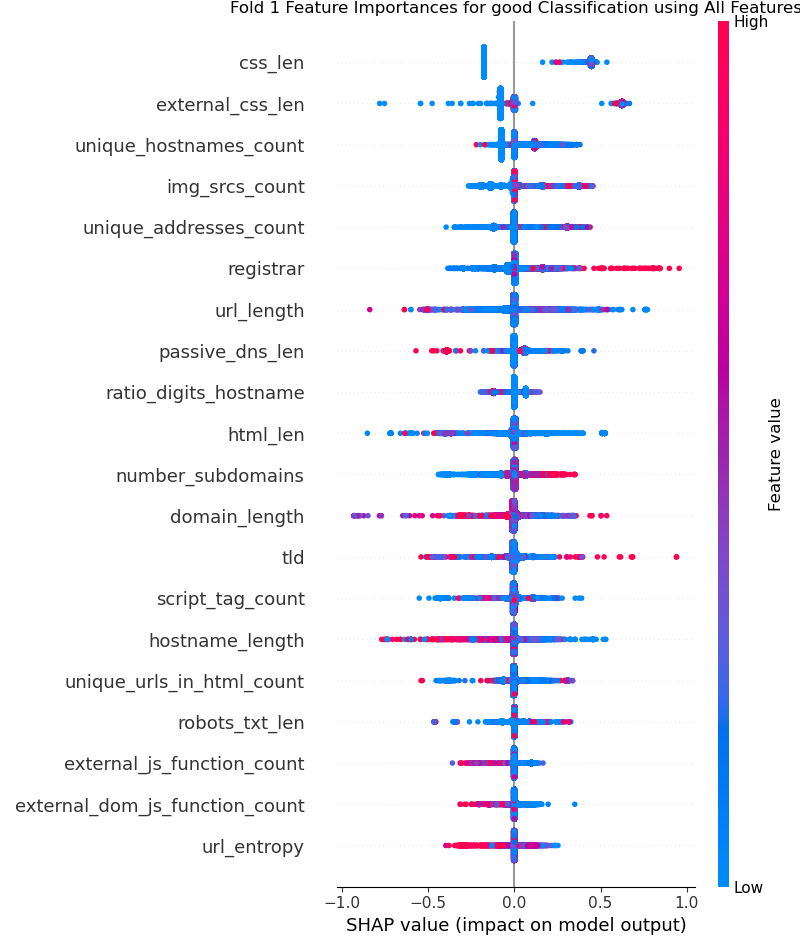} 
    \caption{SHAP Summary Graph of Impact on Random Forest Classifier Output vs Feature Values to the Rank Feature Importance of All Features when making a Benign Classification on the First of Ten Stratified Folds}
    \label{fig:random_forest_feature_rankings_good_all}
\end{figure}

\begin{figure}
    \centering
\includegraphics[width=0.7\textwidth]{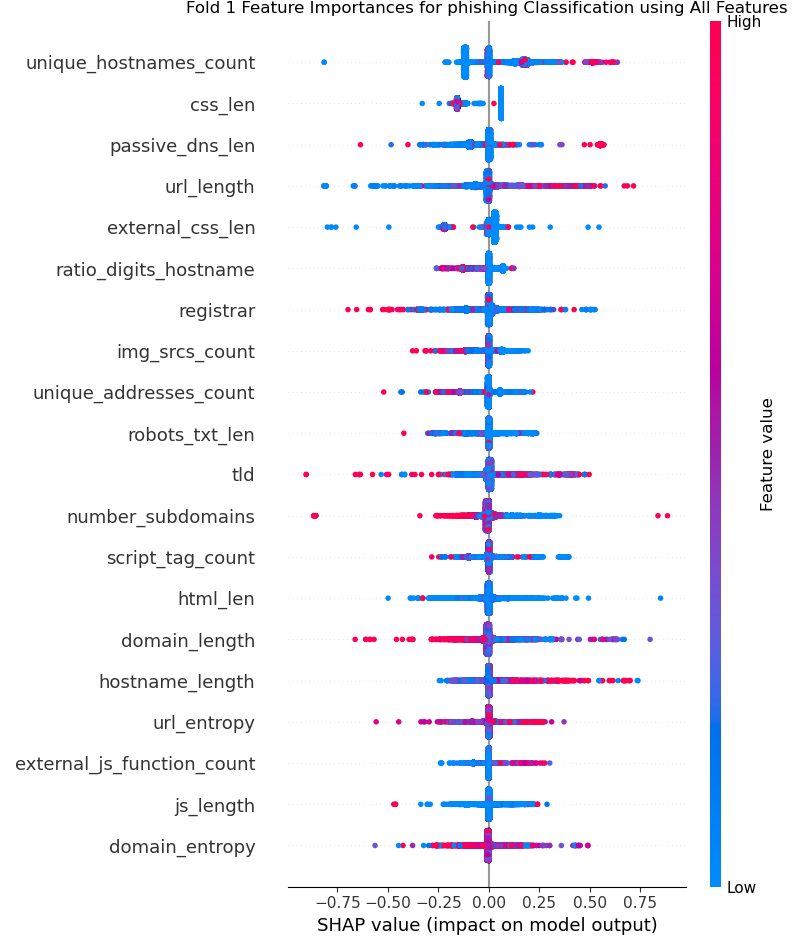} 
    \caption{SHAP Summary Graph of Impact on Random Forest Output vs Feature Values to the Rank Feature Importance of All Features when making a Benign Classification on the First of Ten Stratified Folds}
    \label{fig:random_forest_feature_rankings_phishing_all}
\end{figure}

\begin{table}
    \scriptsize
    \centering
    \caption{The Accuracy Scores of All Models used in Granular Classification Experiments without Hyperparameter Optimizations}
    \vspace{5pt}
    \hspace*{-1cm} 
    \begin{tabular}{lccc}
        \hline
         Accuracy Scores for All Features&  Logistic Regression&  Random Forest Classifier& XGBoost\\
         \hline
         Highest Accuracy&  0.7813&  0.9305& 0.9449\\
         Mean Accuracy&  0.7766 ± 0.0021&  0.928 ± 0.0016& 0.9429 ± 0.001\\
         Highest Top 2 Accuracy&  0.9005&  0.9724& 0.9824 ± 0.0007\\
         Mean Top 2 Accuracy&  0.8988 ± 0.0014&  0.9714 ± 0.0009& 0.974 ± 0.0004\\
         Highest Top 3 Accuracy& 0.9551& 0.9883&0.9944\\
         Mean Top 3 Accuracy& 0.954 ± 0.001& 0.9876 ± 0.0003&0.994 ± 0.0004\\
         \hline
    \end{tabular}
    \label{tab:all_accuracy_scores_no_optimizations}
\end{table}

\begin{table}
    \centering
    \scriptsize
    \caption{Values Assigned to XGBoost Hyperparameters for Training}
    \vspace{5pt}
    \begin{tabular}{lc}
        \hline
         Parameter Name& Assigned Value\\
         \hline
         Max Depth& 7\\
         Minimum child weight& 1\\
         Number of estimators& 165\\
         Colsample by tree& 1.0\\
         Learning rate& 0.3\\
         Tree method& Exact\\
         Booster& Dart\\
         Gamma& $10^{-10}$\\
         \hline
    \end{tabular}
    \label{tab:xgboost_hyperparams}
\end{table}

\begin{table}
    \scriptsize
    \centering
    \caption{The Accuracy Scores of the XGBoost Model used in Granular Classification Experiments without Hyperparameter Optimizations vs with Hyperparameter Optimizations}
    \vspace{5pt}
    \begin{tabular}{lcc}
        \hline
         Accuracy Scores for All Features&  Default Parameters&  Optimized Parameters\\
         \hline
         Highest Accuracy&  0.9449&  0.9541\\
         Mean Accuracy&  0.9429 ± 0.001&  0.9525 ± 0.0011\\
         Highest Top 2 Accuracy&  0.9824 ± 0.0007&  0.9878\\
         Mean Top 2 Accuracy&  0.974 ± 0.0004&  0.9863 ± 0.0009\\
         Highest Top 3 Accuracy& 0.9944& 0.9962\\
         Mean Top 3 Accuracy& 0.994 ± 0.0004& 0.9957 ± 0.0003\\
         \hline
    \end{tabular}
    \label{tab:xgboost_accuracy_scores_default_vs_optimized}
\end{table}

\begin{table}
    \scriptsize
    \centering
    \caption{The Accuracy Scores of the XGBoost Model used in Granular Classification Experiments without Embeddings, with LDA component embeddings, and with 20 best Chi-squared embeddings using All Features}
    \vspace{5pt}
    \begin{tabular}{lccc}
        \hline
         Accuracy Scores for All Features&  No embeddings&  LDA&Chi-squared\\
         \hline
         Highest Accuracy&  0.9541 &  0.9589&0.9576\\
         Mean Accuracy&  0.9525 ± 0.0011 &  0.9568 ± 0.0010&0.9562 ± 0.0008\\
         Highest Top 2 Accuracy&  0.9878 &  0.99&0.9896\\
         Mean Top 2 Accuracy&  0.9863 ± 0.0009 &  0.9885 ± 0.0006&0.988 ± 0.0006\\
         Highest Top 3 Accuracy& 0.9962 & 0.997&0.9962\\
         Mean Top 3 Accuracy& 0.9957 ± 0.0003 & 0.9965 ± 0.0003&0.9958 ± 0.0002\\
         \hline
    \end{tabular}
    \label{tab:xgboost_accuracy_scores_no_embs_vs_lda_vs_chi2}
\end{table}

\begin{table}
    \centering
    \scriptsize
    \caption{Top 10 Contributing Features using All Features (without embeddings)}
    \vspace{5pt}
    \begin{tabular}{lc}
        \hline
        Feature& Contribution (\%)\\
        \hline
        CSS length& 16.2574\\
        Ratio of digits to domain& 10.4236\\
        External CSS length& 8.0536\\
        Unique Hostnames Count& 5.2192\\
        Registrar& 2.5031\\
        Number of equal signs in the URL& 2.3549\\
        Domain length& 2.3006\\
        Number of URLs in the HTML& 2.272\\
        Number of img sources in the HTML& 2.2097\\
        Passive DNS history length&2.1315\\
        \hline
    \end{tabular}
    \label{tab:feature_contributions_no_embs}
\end{table}

\begin{table}
    \centering
    \scriptsize 
    \caption{Top 10 Contributing Features using All Features (with LDA component embeddings)}
    \vspace{5pt}
    \begin{tabular}{lc}
        \hline
        Feature& Contribution (\%)\\
        \hline
        URL (Distilbert) LDA embedding 1& 21.0151\\
        URL (Distilbert) LDA embedding 2& 5.8305\\
        Registrar& 3.2658\\
        Unique Hostnames count& 2.9851\\
        Sitemap count& 2.9836\\
        CSS length& 2.7802\\
        Passive DNS history length& 2.3809\\
        Ratio of digits to characters in the URL& 2.0815\\
        External CSS length&2.066\\
        External JavaScript length&2.03\\
        \hline
    \end{tabular}
    \label{tab:feature_contributions_lda_embs}
\end{table}

\begin{table}
    \centering
    \scriptsize
    \caption{Top 10 Contributing Features using All Features (with 20 best Chi-squared embeddings)}
    \vspace{5pt}
    \begin{tabular}{lc}
        \hline
        Feature& Contribution (\%)\\
        \hline
        CSS length& 16.0121\\
        Ratio of digits to domain& 11.6273\\
        External CSS length& 7.4546\\
        Unique Hostnames Count& 4.7512\\
        Ratio of numbers to characters in the URL& 2.6931\\
        Registrar& 2.5426\\
        Number of URLs in the HTML& 2.5460\\
        Number of img sources in the HTML& 2.2378\\
        Domain length& 2.1717\\
        Passive DNS history length&2.141\\
        \hline
    \end{tabular}
    \label{tab:feature_contributions_chi2_embs}
\end{table}

However, the feature importance rankings did change and include some of the newly introduced features each time a new subset of features was added as shown by comparing Figure \ref{fig:random_forest_feature_rankings_base} and Figure \ref{fig:random_forest_feature_rankings_all}. This suggests that the increasing cascades have an impact on how the logistic regression and random forest models make their benign and malicious classifications. We postulate that the reason for the lower scoring metrics may be due to poorly optimized hyperparameter values since all but the maximum number of features were set to the default. In addition to cascades impacting the models, some of the feature importance rankings from each cascade included features that were proposed in this study and not present in the investigation conducted by Chaiban et al. \cite{app12062806}. For example, in Figure \ref{fig:random_forest_feature_rankings_all} the length of the characters found in the CSS styles of the HTML content was ranked first in importance when considering all features and making a benign classification. This indicated that attempting to develop a larger, more holistic set of features for malicious website classifications was worth pursuing.

\subsection{Granular Classification Experiments}
 From comparing the accuracy scores found in Table \ref{tab:granular_scoring_metrics}, we noticed that this time, as the amount of features increased, the accuracy also increased. Additionally, from comparing Figure \ref{fig:random_forest_feature_rankings_good_base} with Figure \ref{fig:random_forest_feature_rankings_good_all}, we found the feature importance rankings did change to capture the newly cascaded features. This indicates that our methodology is suitable for training models to make granular classifications of malicious websites without making many adjustments to the one used by Chaiban et al \cite{app12062806}. Furthermore, we found that, just like with the previous experiments, the models ranked some of our proposed features highest in feature importance. Figure \ref{fig:random_forest_feature_rankings_good_all} shows that our proposed features: length of CSS, length of external CSS, and registrar made it to the top 1, 2, and 5 features respectively when considering all proposed features. This indicates that our proposed features add necessary details required for making more accurate granular classifications of malicious websites. Lastly, we found that the feature importance rankings were different when it came to making different classifications. This can be observed by comparing Figure \ref{fig:random_forest_feature_rankings_good_all} and Figure \ref{fig:random_forest_feature_rankings_phishing_all}. We then proceeded by training the XGBoost model on our data as it was the best performing model in the study by Chaiban et al \cite{app12062806}.

\subsection{Performance Optimization}
Similar to the investigation performed by Chaiban et al. \cite{app12062806}, the accuracy of the XGBoost model outperformed the rest of the models that we had previously tested. The comparison between all of the model accuracy scores thus far can be found in Table \ref{tab:all_accuracy_scores_no_optimizations}. We followed up on these results by tuning the hyperparameters of the XGBoost model. The chosen hyperparameter values can be found in Table \ref{tab:xgboost_hyperparams}. A comparison of accuracy scores between the hyperparameter tuned XGBoost model and the default XGBoost model can be found in Table \ref{tab:xgboost_accuracy_scores_default_vs_optimized}. 

\subsection{URL Embeddings}
From inspecting Table \ref{tab:xgboost_accuracy_scores_no_embs_vs_lda_vs_chi2}, we learned that the model trained on URL embeddings reduced using linear discriminant analysis had the highest accuracy for all recorded accuracy scores at 95.89\% top 1 accuracy, 0.99\% . This finding is similar to that from Chaiban et al., as their most accurate model was the one that leveraged URL embeddings with PCA reduction. Their model achieved an 84.27\% accuracy compared to our 95.89\%. While this can be attributed to the approximately 50\% increase in features, there are likely other factors that contributed to this difference.  We propose that this increase in accuracy is partially due to the difference in the amount of samples between our studies which lead to the improved performance of our model. Their study used a dataset of 105,485 samples with only 61,080 being benign while our study contained 441,701 samples with 235,721 being benign \cite{app12062806}. 

\subsection{Final Model Feature Contributions}
We calculated the contributions for our three final model candidates. The first candidate contains no embeddings, the second candidate contains URL embeddings that were reduced using LDA, and the last candidate contains URL embeddings that were reduced using Chi-squared feature selection. In the study by Chaiban et al., they noted that their final models did not rank content-based features highly \cite{app12062806}. However, we noticed that was not the case with our models as demonstrated in Table \ref{tab:feature_contributions_no_embs}, Table \ref{tab:feature_contributions_lda_embs}, and Table \ref{tab:feature_contributions_chi2_embs}. Most of the content-based features in our final models were not present in the study by Chaiban et al., which confirms to us that content-based features are important when classifying websites as long as the correct ones are being considered. Additionally, passive DNS history had a considerable contribution to all of our final models and the number sitemaps present in the robots.txt file had a considerable contribution to our highest performing model which shows that a more holistic view of the web should be considered when proposing features. The experimental results and figures can be found on GitHub.\footnote{\url{https://github.com/CyberScienceLab/Malicious_benign_websites}}

\section{Conclusion and Future Work}
\label{sec:conclusion}
In this study, we have expanded upon the work reported by Chaiban et al., to provide an even more holistic dataset with approximately 400\% the number of samples and 150\% the number of features. Most importantly, we have refined the single malicious label by replacing it with 8 unique granular labels. From our experiments, we found that the addition of our proposed features had an impact on model accuracy and feature considerations. Our XGBoost model with LDA reduced URL embeddings achieved a 95.89\% accuracy score which is approximately 11\% increase in accuracy compared to the results of Chaiban et al with their XGBoost model with PCA reduced embeddings \cite{app12062806}. Approximately half of our proposed features consistently ranked in the top 10 contributing features in our 3 best performing models. However, we found that URL embeddings still have the highest feature contribution of all our features. Our results also disagreed with the findings of Chaiban et al., in that we found content-based features were relevant in our most accurate models. The features of high importance in our model were not present in the study by Chaiban et al.  In the future, we would like to see more work exploring the impacts of additional feature categories for classifying benign and malicious websites with our labels. Our work neglected to include networking-based features, or any time-series related features. Such features could lead to more accurate classifications of some malicious labels such as command and control server or malicious advertisement hosting. Future work could also look into the performance of leveraging deep learning neural networks on all features against our XGBoost model.

\section*{Acknowledgment} I would like to extend my deepest gratitude to \textbf{Dr. Ali Dehghantanha} from the University of Guelph, Guelph, Ontario, Canada, for his invaluable guidance and supervision of this work. My sincere thanks also go to \textbf{Stephen Avsec} from Arctic Wolf for stepping in on short notice and providing consistent and invaluable feedback throughout the process. Lastly, I would like to thank \textbf{Dr. Abbas Yazdinejad} from the University of Guelph for his valuable assistance in refining and editing this paper.

\section*{Data Availability}
All scripts, Jupyter notebooks, figures, and references to datasets can be found on our publicly available GitHub repository at \url{https://github.com/CyberScienceLab/Malicious_benign_websites}.
\clearpage

\bibliographystyle{IEEEtran}
\bibliography{main.bib}

\begin{thebibliography}{10}
\providecommand{\url}[1]{#1}
\csname url@samestyle\endcsname
\providecommand{\newblock}{\relax}
\providecommand{\bibinfo}[2]{#2}
\providecommand{\BIBentrySTDinterwordspacing}{\spaceskip=0pt\relax}
\providecommand{\BIBentryALTinterwordstretchfactor}{4}
\providecommand{\BIBentryALTinterwordspacing}{\spaceskip=\fontdimen2\font plus
\BIBentryALTinterwordstretchfactor\fontdimen3\font minus
  \fontdimen4\font\relax}
\providecommand{\BIBforeignlanguage}[2]{{%
\expandafter\ifx\csname l@#1\endcsname\relax
\typeout{** WARNING: IEEEtran.bst: No hyphenation pattern has been}%
\typeout{** loaded for the language `#1'. Using the pattern for}%
\typeout{** the default language instead.}%
\else
\language=\csname l@#1\endcsname
\fi
#2}}
\providecommand{\BIBdecl}{\relax}
\BIBdecl

\bibitem{datareportal2024internet}
\BIBentryALTinterwordspacing
DataReportal, Meltwater, and W.~A. Social, ``Number of internet and social
  media users worldwide as of april 2024 (in billions) [graph],'' Online, 04
  2024, accessed: 2024-08-21. [Online]. Available:
  \url{https://www.statista.com/statistics/617136/digital-population-worldwide/}
\BIBentrySTDinterwordspacing

\bibitem{2}
V.~Santos, T.~Augusto, J.~Vieira, L.~Bacalhau, B.~M. Sousa, and D.~Pontes,
  ``E-commerce: issues, opportunities, challenges, and trends,''
  \emph{Promoting organizational performance through 5G and agile marketing},
  pp. 224--244, 2023.

\bibitem{IC3_FBI_2024}
\BIBentryALTinterwordspacing
{IC3} and {FBI}, ``Annual amount of monetary damage caused by reported
  cybercrime in the united states from 2001 to 2023 (in million u.s. dollars)
  [graph],'' 2024, accessed: 2024-08-21. [Online]. Available:
  \url{https://www.statista.com/statistics/267132/total-damage-caused-by-by-cybercrime-in-the-us/}
\BIBentrySTDinterwordspacing

\bibitem{a2}
A.~Yazdinejad, A.~Dehghantanha, H.~Karimipour, G.~Srivastava, and R.~M. Parizi,
  ``A robust privacy-preserving federated learning model against model
  poisoning attacks,'' \emph{IEEE Transactions on Information Forensics and
  Security}, 2024.

\bibitem{4}
M.~Levi, ``Organising and controlling payment card fraud: Fraudsters and their
  operational environment,'' \emph{Security Journal}, vol.~16, pp. 21--30,
  2003.

\bibitem{app12062806}
\BIBentryALTinterwordspacing
A.~Chaiban, D.~Sovilj, H.~Soliman, G.~Salmon, and X.~Lin, ``Investigating the
  influence of feature sources for malicious website detection,'' \emph{Applied
  Sciences}, vol.~12, no.~6, 2022. [Online]. Available:
  \url{https://www.mdpi.com/2076-3417/12/6/2806}
\BIBentrySTDinterwordspacing

\bibitem{8780783}
B.~Chen and Y.~Shi, ``Malicious hidden redirect attack web page detection based
  on css features,'' in \emph{2018 IEEE 4th International Conference on
  Computer and Communications (ICCC)}, 2018, pp. 1155--1159.

\bibitem{10.1145/3387168.3387236}
\BIBentryALTinterwordspacing
Z.~Bao, W.~Wang, and Y.~Lan, ``Using passive dns to detect malicious domain
  name,'' ser. ICVISP 2019.\hskip 1em plus 0.5em minus 0.4em\relax New York,
  NY, USA: Association for Computing Machinery, 2020. [Online]. Available:
  \url{https://doi.org/10.1145/3387168.3387236}
\BIBentrySTDinterwordspacing

\bibitem{ibm_incidence_response}
\BIBentryALTinterwordspacing
IBM, ``What is incident response?'' 2024, accessed: 2024-08-21. [Online].
  Available: \url{https://www.ibm.com/topics/incident-response}
\BIBentrySTDinterwordspacing

\bibitem{me}
\BIBentryALTinterwordspacing
K.~Tran, ``Malicious benign websites,'' 08 2024, accessed: 2024-08-26.
  [Online]. Available:
  \url{https://github.com/CyberScienceLab/Malicious_benign_websites}
\BIBentrySTDinterwordspacing

\bibitem{a1}
A.~Yazdinejad, R.~M. Parizi, A.~Dehghantanha, Q.~Zhang, and K.-K.~R. Choo, ``An
  energy-efficient sdn controller architecture for iot networks with
  blockchain-based security,'' \emph{IEEE Transactions on Services Computing},
  vol.~13, no.~4, pp. 625--638, 2020.

\bibitem{SINGH2020106304}
\BIBentryALTinterwordspacing
A.~Singh, ``Malicious and benign webpages dataset,'' \emph{Data in Brief},
  vol.~32, p. 106304, 2020. [Online]. Available:
  \url{https://www.sciencedirect.com/science/article/pii/S2352340920311987}
\BIBentrySTDinterwordspacing

\bibitem{10.1007/978-3-319-50472-8_17}
A.~K. Singh and N.~Goyal, ``Malcrawler: A crawler for seeking and crawling
  malicious websites,'' in \emph{Distributed Computing and Internet
  Technology}, P.~Krishnan, P.~Radha~Krishna, and L.~Parida, Eds.\hskip 1em
  plus 0.5em minus 0.4em\relax Cham: Springer International Publishing, 2017,
  pp. 210--223.

\bibitem{kaggle}
\BIBentryALTinterwordspacing
M.~Siddhartha, ``Malicious urls dataset,'' 07 2021, accessed: 2024-06-30.
  [Online]. Available:
  \url{https://www.kaggle.com/datasets/sid321axn/malicious-urls-dataset}
\BIBentrySTDinterwordspacing

\bibitem{s22093373}
\BIBentryALTinterwordspacing
M.~Alsaedi, F.~A. Ghaleb, F.~Saeed, J.~Ahmad, and M.~Alasli, ``Cyber threat
  intelligence-based malicious url detection model using ensemble learning,''
  \emph{Sensors}, vol.~22, no.~9, 2022. [Online]. Available:
  \url{https://www.mdpi.com/1424-8220/22/9/3373}
\BIBentrySTDinterwordspacing

\bibitem{https://doi.org/10.1155/2022/3241216}
\BIBentryALTinterwordspacing
M.~Aljabri, F.~Alhaidari, R.~M.~A. Mohammad, S.~Mirza, D.~H. Alhamed, H.~S.
  Altamimi, and S.~M.~B. Chrouf, ``An assessment of lexical, network, and
  content-based features for detecting malicious urls using machine learning
  and deep learning models,'' \emph{Computational Intelligence and
  Neuroscience}, vol. 2022, no.~1, p. 3241216. [Online]. Available:
  \url{https://onlinelibrary.wiley.com/doi/abs/10.1155/2022/3241216}
\BIBentrySTDinterwordspacing

\bibitem{MCGAHAGAN2021102374}
\BIBentryALTinterwordspacing
J.~McGahagan, D.~Bhansali, C.~Pinto-Coelho, and M.~Cukier, ``Discovering
  features for detecting malicious websites: An empirical study,''
  \emph{Computers \& Security}, vol. 109, p. 102374, 2021. [Online]. Available:
  \url{https://www.sciencedirect.com/science/article/pii/S016740482100198X}
\BIBentrySTDinterwordspacing

\bibitem{naim2023malicious}
\BIBentryALTinterwordspacing
O.~Naim, D.~Cohen, and I.~Ben-Gal, ``Malicious website identification using
  design attribute learning,'' \emph{International Journal of Information
  Security}, vol.~22, no.~5, pp. 1207--1217, 2023. [Online]. Available:
  \url{https://doi.org/10.1007/s10207-023-00686-y}
\BIBentrySTDinterwordspacing

\bibitem{beautiful_soup}
\BIBentryALTinterwordspacing
L.~Richardson, ``Beautiful soup documentation,'' 2024, accessed: 2024-05-21.
  [Online]. Available:
  \url{https://www.crummy.com/software/BeautifulSoup/bs4/doc/}
\BIBentrySTDinterwordspacing

\bibitem{tldextract}
\BIBentryALTinterwordspacing
J.~Kurkowski, ``tldextract 5.1.2,'' 2024, accessed: 2024-05-21. [Online].
  Available: \url{https://pypi.org/project/tldextract/}
\BIBentrySTDinterwordspacing

\bibitem{urllib3}
\BIBentryALTinterwordspacing
A.~Petrov, ``urllib3,'' 2024, accessed: 2024-05-21. [Online]. Available:
  \url{https://urllib3.readthedocs.io/en/stable/}
\BIBentrySTDinterwordspacing

\bibitem{distilbert}
\BIBentryALTinterwordspacing
V.~Sanh, ``Distilbert,'' 2024, accessed: 2024-08-09. [Online]. Available:
  \url{https://huggingface.co/docs/transformers/v4.44.1/en/model_doc/distilbert#overview}
\BIBentrySTDinterwordspacing

\bibitem{longformer}
\BIBentryALTinterwordspacing
I.~Beltagy, ``Longformer,'' 2024, accessed: 2024-08-09. [Online]. Available:
  \url{https://huggingface.co/docs/transformers/en/model_doc/longformer}
\BIBentrySTDinterwordspacing

\bibitem{IBM_X-force}
\BIBentryALTinterwordspacing
I.~Security, ``X-force threat intelligence api,'' 2024, accessed: 2024-06-30.
  [Online]. Available: \url{https://api.xforce.ibmcloud.com/doc/}
\BIBentrySTDinterwordspacing

\bibitem{alienvault}
\BIBentryALTinterwordspacing
L.~Labs, ``The world’s first truly open threat intelligence community,''
  2024, accessed: 2024-06-30. [Online]. Available:
  \url{https://otx.alienvault.com/}
\BIBentrySTDinterwordspacing

\bibitem{mozilla_dom_api}
\BIBentryALTinterwordspacing
M.~Foundation, ``Document object model (dom),'' 2024, accessed: 2024-06-30.
  [Online]. Available:
  \url{https://developer.mozilla.org/en-US/docs/Web/API/Document_Object_Model}
\BIBentrySTDinterwordspacing

\bibitem{easylist}
\BIBentryALTinterwordspacing
R.~Petnel, ``Easylist,'' 2024, gitHub Repository. Available online:
  https://github.com/easylist/easylist. Accessed: 2024-06-30. [Online].
  Available: \url{https://easylist.to/}
\BIBentrySTDinterwordspacing

\bibitem{ICANN_fee}
\BIBentryALTinterwordspacing
I.~C. for Assigned~Names and Numbers, ``Registrar fees,'' 2024, accessed:
  2024-06-30. [Online]. Available:
  \url{https://www.icann.org/resources/pages/registrar-fees-2018-08-10-en}
\BIBentrySTDinterwordspacing

\bibitem{urlhaus}
\BIBentryALTinterwordspacing
abuse.ch, ``Urlhaus,'' 2024, accessed: 2024-06-30. [Online]. Available:
  \url{https://urlhaus.abuse.ch/}
\BIBentrySTDinterwordspacing

\bibitem{threatfox}
\BIBentryALTinterwordspacing
------, ``Threatfox,'' 2024, accessed: 2024-06-30. [Online]. Available:
  \url{https://threatfox.abuse.ch/}
\BIBentrySTDinterwordspacing

\bibitem{a3}
B.~Zolfaghari, A.~Yazdinejad, A.~Dehghantanha, J.~Krzciok, and K.~Bibak, ``The
  dichotomy of cloud and iot: Cloud-assisted iot from a security perspective,''
  \emph{arXiv preprint arXiv:2207.01590}, 2022.

\bibitem{a4}
K.~Viswanathan and A.~Yazdinejad, ``Security considerations for virtual reality
  systems,'' \emph{arXiv preprint arXiv:2201.02563}, 2022.

\bibitem{linear_discriminant_analysis}
\BIBentryALTinterwordspacing
R.~A. FISHER, ``The use of multiple measurements in taxonomic problems,''
  \emph{Annals of Eugenics}, vol.~7, no.~2, pp. 179--188, 1936. [Online].
  Available:
  \url{https://onlinelibrary.wiley.com/doi/abs/10.1111/j.1469-1809.1936.tb02137.x}
\BIBentrySTDinterwordspacing

\bibitem{jolliffe2002pca}
I.~T. Jolliffe, \emph{Principal Component Analysis}, 2nd~ed., ser. Springer
  Series in Statistics.\hskip 1em plus 0.5em minus 0.4em\relax Springer, 2002.

\end{thebibliography}
\end{document}